\documentclass{article} 
\usepackage{mlgenx2024,times}

\usepackage{amsmath,amsfonts,bm}

\def\eqref#1{equation~\ref{#1}}

\def\1{\bm{1}}

\DeclareMathAlphabet{\mathsfit}{\encodingdefault}{\sfdefault}{m}{sl}
\SetMathAlphabet{\mathsfit}{bold}{\encodingdefault}{\sfdefault}{bx}{n}

\usepackage{hyperref}
\usepackage{url}
\usepackage{booktabs}
\usepackage{amsmath}
\usepackage{amssymb}
\usepackage{mathtools}
\usepackage{amsthm}
\usepackage{enumitem}
\usepackage{multirow}
\usepackage{caption}
\usepackage{subcaption}
\usepackage{capt-of}
\usepackage{wrapfig}
\usepackage{bm}
\usepackage{bbm}
\usepackage{arydshln}
\usepackage{placeins}
\usepackage{comment}
\usepackage{algorithm}
\usepackage{algorithmic}
\usepackage{eqparbox}
\usepackage{colortbl}
\usepackage{tikz}
\usepackage{graphicx}
\usepackage{tabularx}
\usepackage{adjustbox}
\usepackage{wrapfig}
\usepackage{caption}
\usepackage{colortbl}

\title{Protein Representation Learning \\by Capturing Protein \\Sequence-Structure-Function Relationship}

\author{Eunji Ko$^{1*}$ \quad Seul Lee$^{1*}$ \quad Minseon Kim$^{1}$\thanks{Equal contribution} \quad Dongki Kim$^{1}$ \quad Sung Ju Hwang$^{1}$ \\
$^{1}$KAIST, South Korea \\
\texttt{\{kosu7071,seul.lee,minseonkim,cleverki,sjhwang82\}@kaist.ac.kr} \\
}

\iclrfinalcopy 
\begin{document}

\maketitle

\begin{abstract}
The goal of protein representation learning is to extract knowledge from protein databases that can be applied to various protein-related downstream tasks. Although protein sequence, structure, and function are the three key modalities for a comprehensive understanding of proteins, existing methods for protein representation learning have utilized only one or two of these modalities due to the difficulty of capturing the asymmetric interrelationships between them. To account for this asymmetry, we introduce our novel \emph{asymmetric multi-modal masked autoencoder} (AMMA). AMMA adopts (1) a unified multi-modal encoder to integrate all three modalities into a unified representation space and (2) asymmetric decoders to ensure that sequence latent features reflect structural and functional information. The experiments demonstrate that the proposed AMMA is highly effective in learning protein representations that exhibit well-aligned inter-modal relationships, which in turn makes it effective for various downstream protein-related tasks.
\end{abstract}

\vspace{-0.1in}
\section{Introduction}

\vspace{-0.1in}

Proteins are generated in an organism in the form of a sequence, which is then folded into a three-dimensional structure, and as a three-dimensional structure, they become functional and fulfill their roles. This is the so-called protein sequence-structure-function paradigm~\citep{liberles2012interface,serccinouglu2020sequence}. Of the three modalities—sequence, structure, and function—sequence information underlies many protein applications and is the most abundant, making it a popular choice for training neural networks. The challenge lies in developing sophisticated protein representations that utilize information across various modalities based on sequence data. However, existing methods for protein representation learning have only utilized some of the modalities, overlooking the importance of comprehensive integration of these modalities.

\begin{figure*}[t]
    \centering
    \begin{minipage}{0.68\textwidth}
        \centering
        \includegraphics[width=0.95\linewidth]{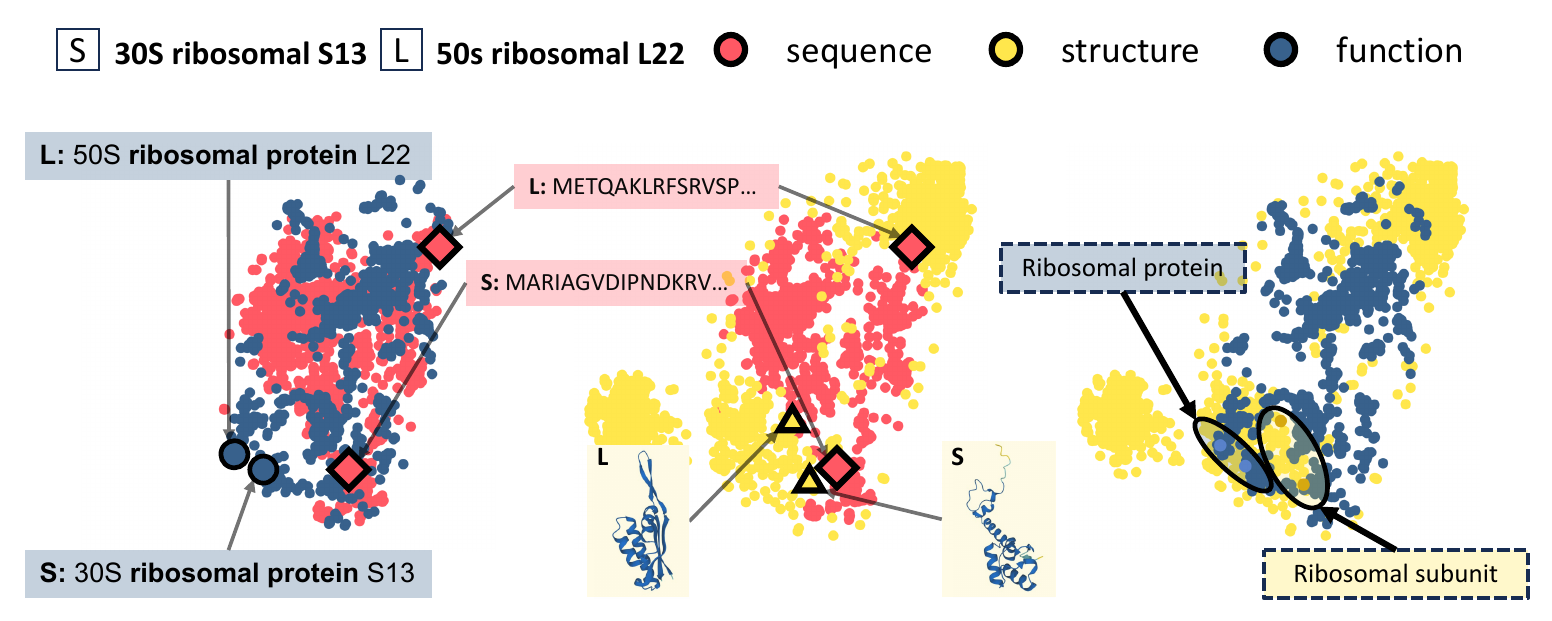}
        \vspace{-0.1in}
        \caption{\small \textbf{t-SNE visualization of the three modalities of proteins.} Latent features for sequence ({\color{red}red}), structure ({\color{yellow}yellow}), and function ({\color{blue}blue}) extracted from ESM-1b, GearNet, and PubMedBERT-abs, respectively, are visualized. Two proteins, 30S ribosomal protein S13 (S) and 50S ribosomal protein L22 (L), are functionally similar and therefore proximal in function space (left). These proteins are encoded close together in structure space, but far apart in sequence space (middle). This trend is common across ribosomal proteins (right). Details are provided in Section~\ref{sec:tsne_detail}.}
        \label{fig:tsne}
    \end{minipage}
    \hfill
    \begin{minipage}{0.3\textwidth}
        \vspace{0.1in}
        \captionof{table}{\small \textbf{Cosine similarity scores} between the relation matrices which calculate relationship between the protein latents in a batch. The latents are extracted from the uni-modal encoders (i.e., ESM-1b, GearNet, and PubMedBERT-abs with additional projection layers). We report the similarity values calculated before and after applying contrastive learning (CL) and our proposed AMMA. Details are provided in Section~\ref{sec:csim_detail}.}
        \vspace{-0.1in}
        \begin{adjustbox}{width=\linewidth}{
            \begin{tabular}{c|ccc}
                \toprule
                Alignment & Initial & CL & AMMA \\
                \midrule
                Seq-Str & 0.865 & 0.894 & 1.000 \\
                Seq-Func & 0.855 & 0.907 & 0.999 \\
                Str-Func & 0.947 & 0.847 & 0.999 \\
                \bottomrule
            \end{tabular}
        }
        \end{adjustbox}
        \label{tab:similarity}
    \end{minipage}
    \vspace{-0.25in}
\end{figure*}

A significant hurdle to comprehensively considering the three modalities is the complexity of capturing the relationship between them. The correspondence between them is not straightforward, for example, even if the amino acid sequences are very similar, substrate specificity can change dramatically as the three-dimensional structure of the active site changes~\citep{bunsupa2012lysine}. Moreover, proteins that have acquired the same function by convergent evolution, or that have accumulated sequence mutations where they do not affect protein folding, can have very little similarity in sequence. As has been described by many literatures~\citep{illergaard2009structure,mahlich2018hfsp,van2022foldseek}, it is a generally accepted fact that it is the structure, not the sequence, that is more conserved and directly related to the function of a protein. Figure~\ref{fig:tsne} empirically supports this claim, showing that proteins with similar functional features are encoded into similar structural features, while their sequence features can differ largely. The first column of Table~\ref{tab:similarity} quantitatively shows that the structure-function relationship is correlated more compared to the sequence-structure or sequence-function relationships. We refer to this relationship, where structure and function exhibit a strong alignment, while sequence and the other two show a relatively weaker correlation, as the \emph{asymmetric relationship} between modalities.

To utilize information from multiple modalities, albeit not all three, most previous works~\citep{zhang2022ontoprotein,xu2023protst,zhang2024pepharmony} have leveraged contrastive learning, which learns the instance similarity and difference between two modalities. However, these approaches focus on improving representations of a single modality (e.g., protein sequence) by utilizing guidance from other ``auxiliary'' modalities. This leads to a skewed integration of modalities, as shown in the second column of Table~\ref{tab:similarity}, where contrastive learning shows high sequence-structure and sequence-function similarity but low structure-function similarity.

To this end, we propose \emph{Asymmetric Multi-modal Masked Autoencoder} (AMMA), an integrated protein representation learning method that jointly embeds the three core modalities of proteins. Under the masked autoencoder framework~\citep{bachmann2022multimae}, AMMA captures the asymmetric sequence-structure-function relationship inherent in the protein domain through (1) the unified multi-modal encoder and (2) the asymmetric design of the decoder. By having the multi-modal encoder, AMMA explicitly integrates information from all of the modalities rather than letting one modality guide the others. Furthermore, by learning to predict structure and function from sequence with the asymmetric decoders, AMMA ensures that sequence latent features faithfully reflect structure and function information. As shown in the last column of Table~\ref{tab:similarity}, AMMA successfully yields well-aligned multi-modal protein representations. We experimentally validate the proposed AMMA on various tasks that require accurate protein representation learning. The experimental results demonstrate that AMMA outperforms existing state-of-the-art methods, showing its superiority in learning protein representations by comprehensively and effectively considering the multi-modal aspects of proteins. We summarize our contributions as follows:
\begin{itemize}
    \vspace{-0.1in}
    \item We are the first to propose utilizing the three core modalities for protein representation learning: sequence, structure, and function.
    \item We point out the asymmetric relationship between sequence, structure, and function of proteins and propose AMMA, a masked autoencoder framework that adopts a unified multi-modal encoder and asymmetric decoders to account for the asymmetric relationship.
    \item We experimentally demonstrate that AMMA is highly effective in learning protein representations and benefits performance on a variety of downstream protein-related tasks.
\end{itemize}
\vspace{-0.2in}
\section{Related works}
\vspace{-0.1in}
As a means to overcome the limitations of uni-modal protein representation learning, protein representation learning using multiple modalities has gained traction. Most previous studies have adopted a contrastive learning approach to capture the relationship between modalities. \citet{zhang2022ontoprotein} employed knowledge-aware negative sampling to identify negative instances, enabling contrastive learning across proteins. \citet{xu2023protst} conducted contrastive learning between protein sequence and functional description. However, contrastive learning may not be an optimal for multi-modal representation learning, as it focuses on learning improved representations of a single modality using other modal information and thus cannot yield balanced multi-modal representations.

Apart from contrastive learning, there are other approaches to multi-modal protein representation learning. \citet{su2023saprot} proposed using structure-aware tokens from FoldSeek~\citep{van2022foldseek} to train a token-based ESM~\citep{lin2022language}. However, the method of \citet{su2023saprot} uses 20 different structural tokens, which restricts diversity when encoding structural information. Moreover, these methods do not consider another important modality for proteins, the function description, and are therefore suboptimal. 
\begin{figure*}[t]
    \centering
    \includegraphics[width=0.9\textwidth]{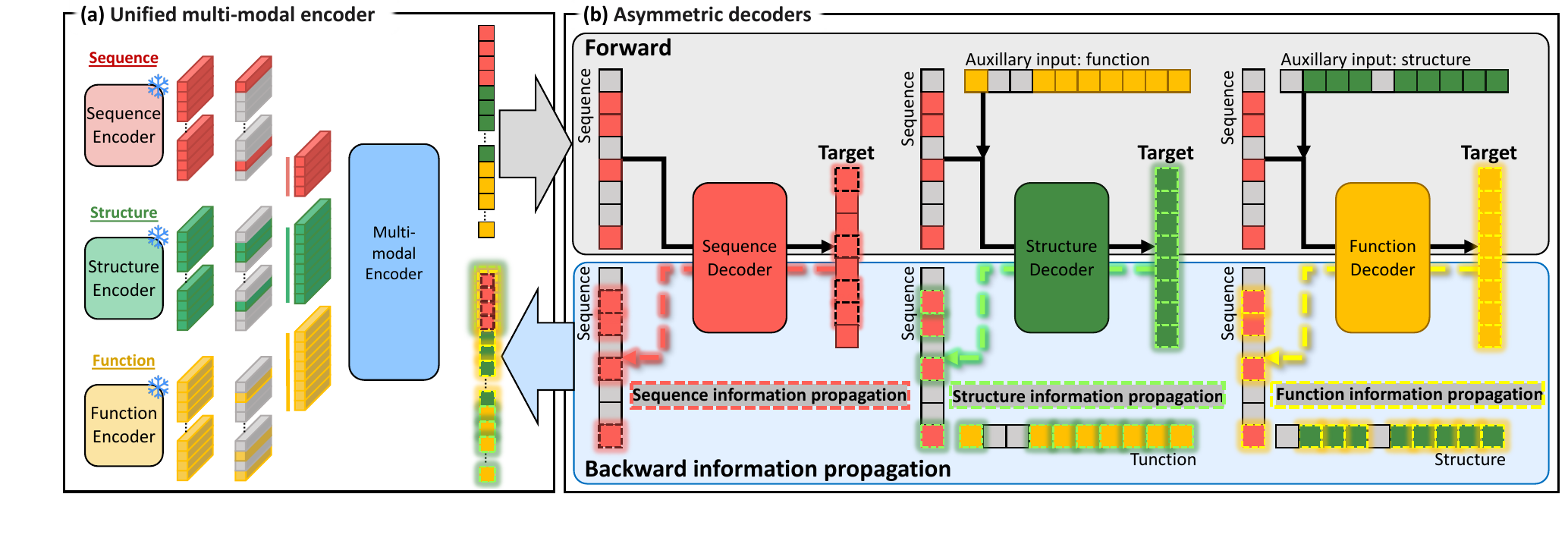}
    \vspace{-0.16in}
    \caption{\small \textbf{Multi-modal protein representation learning with AMMA.} AMMA has two key components: (a) a unified multi-modal encoder and (b) asymmetric decoders. Each modality is encoded by a frozen pretrained encoder, then integrated by a multi-modal encoder after masking. Asymmetric decoders then reconstruct original features of each modality. During decoding, the input latent features, designed to hold target-specific information, are asymmetrically passed to the decoders for target modality reconstruction. This requires AMMA to encode structural and functional information into sequence latent features, which allows AMMA to capture unique asymmetric sequence-structure-function relationships. The overall architecture is provided in Figure~\ref{fig:amma}.}
    \label{fig:concept}
    \vspace{-0.22in}
\end{figure*}
\vspace{-0.1in}
\section{Method}
\label{sec:method}




\vspace{-0.1in}
\subsection{Asymmetric Multi-modal Masked Autoencoder (AMMA)}
\label{sec:framework}
\vspace{-0.1in}
\paragraph{Sequence, Structure, and Function Uni-modal Encoders~\label{sec:encoder}}
To integrate the protein information of sequence, structure, and function equally to construct a unified protein representation, we first propose to utilize a single multi-modal encoder. The inputs to the multi-modal encoder are the features extracted from each of the uni-modal encoders. The input data can be the sequence $X_{\texttt{seq}}$, structure $X_{\texttt{str}}$, or function $X_{\texttt{func}}$, or a combination of these modalities. We adopt pretrained feature extractors as the uni-modal encoders $\textsc{Enc}_\texttt{seq}$, $\textsc{Enc}_\texttt{str}$, and $\textsc{Enc}_\texttt{func}$. Specifically, we use ESM-1b~\citep{rives2021biological}, GearNet~\citep{zhang2022protein}, and PubMedBERT-abs~\citep{gu2021domain} for extracting features from the sequence, structure, and function inputs, respectively. Note that our approach is model-agnostic, and any off-the-shelf uni-modal feature extractors can be used. $X_{\texttt{seq}}$, $X_{\texttt{str}}$, and $X_{\texttt{func}}$ are each passed to the corresponding uni-modal encoder and becomes $X_{\texttt{seq}}' \in \mathbb{R}^{L\times1280}$, $X_{\texttt{str}}' \in \mathbb{R}^{L\times3072}$, and $X_{\texttt{func}}' \in \mathbb{R}^{L'\times768}$.
$L$ is the number of amino acids of the protein and $L'$ is the number of the tokens of the function description. Then, we add two fully connected layers, $\textsc{Proj}_{\texttt{seq}}$, $\textsc{Proj}_{\texttt{str}}$, and $\textsc{Proj}_{\texttt{func}}$, to each of the uni-modal encoders to project the latent features of multiple modalities to the same size of projection dimension. The result becomes $Z_{\texttt{seq}} \in \mathbb{R}^{L\times D}$, $Z_{\texttt{str}} \in \mathbb{R}^{L\times D}$, and $Z_{\texttt{func}} \in \mathbb{R}^{L'\times D}$. 
$D$ is the projection dimension. We set $D$ to 512.

\vspace{-0.1in}
\paragraph{Mask Sampling~\label{sec:masking}}
Under the masked autoencoder framework, we mask the encoded latent features $Z$. Specifically, we first sample the preserving ratios between the modalities, $\lambda_{\texttt{seq}}$, $\lambda_{\texttt{str}}$, and $\lambda_{\texttt{func}}$, based on the Dirichlet distribution following \citet{bachmann2022multimae}, where $\lambda_{\texttt{seq}} + \lambda_{\texttt{str}} + \lambda_{\texttt{func}} = 1$, $\lambda_{\texttt{seq}} \geq 0$, $\lambda_{\texttt{str}} \geq 0$, and $\lambda_{\texttt{func}} \geq 0$ as follows:
\begin{align}
    (\lambda_{\texttt{seq}}, \lambda_{\texttt{str}}, \lambda_{\texttt{func}}) \sim \text{Dirichlet}(\alpha_{\texttt{seq}}, \alpha_{\texttt{str}}, \alpha_{\texttt{func}}).
\end{align}
We set the value of $\alpha_{\texttt{seq}}$, $\alpha_{\texttt{str}}$, and $\alpha_\texttt{func}$ to 1, 2, and 2, respectively. We then randomly mask the latent features $Z_{\texttt{seq}}$, $Z_{\texttt{str}}$, and $Z_{\texttt{func}}$ such that the number of preserved tokens has the ratio $\lambda_{\texttt{seq}} : \lambda_{\texttt{str}} : \lambda_{\texttt{func}}$ and sums to the total number of tokens $M$. Subsequently, we concatenate the masked features from all three modalities to be $Z\in \mathbb{R}^{M\times D}$, which will be the input to the multi-modal encoder described in the next paragraph. We set $M$ to 160.

\vspace{-0.1in}
\paragraph{Multi-modal Encoder}
To learn protein representations that faithfully contain information from multiple modalities uniformly in a well-aligned manner, we propose to use a unified protein multi-modal encoder $\textsc{Enc}_{\texttt{multi}}$ that integrates different modalities into a single representation space. $\textsc{Enc}_{\texttt{multi}}$ encodes the concatenated and masked multi-modal latent features as follows:
\begin{align}
    Z_\texttt{multi} = \textsc{Enc}_{\texttt{multi}}(Z) \in \mathbb{R}^{M\times D}.
\end{align}
We adopt an 8-layer Transformer as $\textsc{Enc}_{\texttt{multi}}$. Through the self-attention mechanism, $\textsc{Enc}_{\texttt{multi}}$ facilitates the fusion of multiple modality information.

\vspace{-0.07in}
\paragraph{Asymmetric Decoder~\label{sec:decoder}}
To train the multi-modal encoder under the autoencoder framework, we adopt individual modality decoders that reconstruct the original latent variables $X_{\texttt{seq}}'$, $X_{\texttt{str}}'$, and $X_{\texttt{func}}'$, respectively. Unlike multi-modal representation learning in the image domain, multi-modal protein representation learning should consider that the three protein modalities exhibit a unique asymmetric relationship in which sequence-structure and sequence-function are relatively poorly aligned compared to structure-function (see Figure~\ref{fig:tsne}). To effectively represent proteins by incorporating information across modalities and capturing their asymmetric interrelationships, we introduce asymmetric decoders. 

The multi-modal latent variable $Z_\texttt{multi}$ computed by the multi-modal encoder $\textsc{Enc}_\texttt{multi}$ is first passed to each of the three single linear layers $\ell_0$, $\ell_1$, and $\ell_2$ and becomes $Z'_{\texttt{multi},0} \in \mathbb{R}^{M \times D}$, $Z'_{\texttt{multi},1} \in \mathbb{R}^{M \times D}$, and $Z'_{\texttt{multi},2} \in \mathbb{R}^{M \times D}$.
Tokens of zero are then inserted into the masked positions of $Z'_{\texttt{multi},0}$, $Z'_{\texttt{multi},1}$, and $Z'_{\texttt{multi},2}$ to regain the original unmasked length $2L+L'$. The three latent features of size $(2L+L')\times D$ each split into three modal-specific latent features $Z'_{\texttt{seq},k} \in \mathbb{R}^{L \times D}$, $Z'_{\texttt{str},k} \in \mathbb{R}^{L \times D}$, and $Z'_{\texttt{func},k} \in \mathbb{R}^{L' \times D}$ for $k=0,1,2$. Subsequently, the latent features are asymmetrically passed to modal-specific decoders $\textsc{Dec}_\texttt{seq}$, $\textsc{Dec}_\texttt{str}$, and $\textsc{Dec}_\texttt{func}$ depending on the modality as follows:
\vspace{-0.05in}
\begin{align}
\begin{split}
    \hat{X}_{\texttt{seq}} &= \textsc{Dec}_\texttt{seq}(Z'_{\texttt{seq},0}) \in \mathbb{R}^{L\times1280}, \\
    \hat{X}_{\texttt{str}} &= \textsc{Dec}_\texttt{str}(Z'_{\texttt{seq},1}, \, Z'_{\texttt{func},1}) \in \mathbb{R}^{L\times3072}, \\
    \hat{X}_{\texttt{func}} &= \textsc{Dec}_\texttt{func}(Z'_{\texttt{seq},2}, \, Z'_{\texttt{str},2})  \in \mathbb{R}^{L'\times768},
    \label{eq:decoder}
\end{split}
\end{align}
where $\hat{X}_{\texttt{seq}}$, $\hat{X}_{\texttt{str}}$, and $\hat{X}_{\texttt{func}}$ are the reconstructed latent features of $X'_{\texttt{seq}}$, $X'_{\texttt{str}}$, and $X'_{\texttt{func}}$, respectively. Each decoder consists of two Transformer layers followed by a single linear layer. During the pretraining, we keep the uni-modal encoders frozen and train AMMA using the mean squared error (MSE) loss as follows:
\begin{align}
\begin{split}
    \mathcal{L}_{\texttt{seq}} = \text{MSE}(\hat{X}_{\texttt{seq}}, \, X_{\texttt{seq}}'), 
    \mathcal{L}_{\texttt{str}} =& \text{MSE}(\hat{X}_{\texttt{str}}, \, X_{\texttt{str}}'), 
    \mathcal{L}_{\texttt{func}} = \text{MSE}(\hat{X}_{\texttt{func}}, \, X_{\texttt{func}}'), \\
    \mathcal{L} =& \mathcal{L}_{\texttt{seq}} + \mathcal{L}_{\texttt{str}} + \mathcal{L}_{\texttt{func}}.
\end{split}
\end{align}

\vspace{-0.1in}
The key to the proposed AMMA is the specialized asymmetric design of each decoder. The sequence decoder takes $Z'_\texttt{seq}$ as input to predict the original sequence latent feature. This ensures that the sequence information remains intact during the fusion process. On the other hand, the structure decoder takes both $Z'_\texttt{seq}$ and $Z'_\texttt{func}$ as input to reconstruct the original structural latent feature. This design is crucial to ensure the sequence latent features to reflect structural information during the fusion process of the multi-modal encoder. Similarly, the function decoder takes both $Z'_\texttt{seq}$ and $Z'_\texttt{str}$ as input to predict the original function latent feature, ensuring the sequence latent features to reflect function information during the fusion process. Because structure and function are difficult to predict from sequence features alone and structure and function are relatively well aligned, structure and function can provide the auxiliary information needed to reconstruct each other. Using other modalities as auxiliary information for the structure and function decoders is more effective than using their own features as auxiliary information. This strategy prevents the decoders from overly relying on their own explicit inputs, which can weaken the modality information propagating backwards to the sequence features, preventing the integration of the modalities.

Through the proposed multi-modal encoder and asymmetric decoders, AMMA adeptly learns to encapsulate all three core protein modalities: sequence, structure, and function. This results in effective and comprehensive multi-modal protein representations that reflects the complex interdependencies of protein modalities. 
\vspace{-0.1in}
\section{Experiments}

\begin{table*}[t]
    \caption{\small \textbf{Performance on protein function annotation tasks.}} 
    \vspace{-0.1in}
    \centering
    \resizebox{\textwidth}{!}{
    \renewcommand{\tabcolsep}{2.5mm}
    \begin{tabular}{l|ccc|cccccccccc}
    \toprule
    \multirow{2.5}{*}{Method} & \multicolumn{3}{c|}{Modality} & \multicolumn{2}{c}{EC} & \multicolumn{2}{c}{GO-MF} & \multicolumn{2}{c}{GO-CC} & \multicolumn{2}{c}{GO-BP} & \multirow{2.5}{*}{$\text{Avg.}_\text{Fmax}$} & \multirow{2.5}{*}{$\text{Avg.}_\text{AUPR}$} \\
    \cmidrule(l{2pt}r{2pt}){2-4}
    \cmidrule(l{2pt}r{2pt}){5-6}
    \cmidrule(l{2pt}r{2pt}){7-8}
    \cmidrule(l{2pt}r{2pt}){9-10}
    \cmidrule(l{2pt}r{2pt}){11-12}
    & Seq. & Str. & Func. & $\text{F}_\text{max}$ & AUPR & $\text{F}_\text{max}$ & AUPR & $\text{F}_\text{max}$ & AUPR & $\text{F}_\text{max}$ & AUPR \\
    \midrule
    ESM-1b~\citep{rives2021biological} & \checkmark & & & 86.9 & 88.4 & 65.9 & 63.0 & 47.7 & 32.4 & 45.2 & \underline{33.2} & \cellcolor{gray!25} 61.4 & \cellcolor{gray!25} 54.3 \\
    OntoProtein~\citep{zhang2022ontoprotein} & \checkmark & & \checkmark & 84.1 & 85.4 & 63.1 & 60.3 & 44.1 & 30.0 & 43.6 & 28.4 & \cellcolor{gray!25} 58.7 & \cellcolor{gray!25} 51.0 \\
    GearNet~\citep{zhang2022protein} &  & \checkmark & & 87.4 & 89.2 & 65.4 & 59.6 & \underline{48.8} & 33.6 & \textbf{49.0} & 29.2 & \cellcolor{gray!25} \underline{62.7} & \cellcolor{gray!25} 52.9 \\
    SaProt~\citep{su2023saprot} & \checkmark & \checkmark & & \textbf{88.8} & 85.5 & \textbf{68.8} & 58.2 & 41.2 & 20.6 & 45.1 & 23.8  & \cellcolor{gray!25} 61.0 & \cellcolor{gray!25} 47.0 \\
    ProtST~\citep{xu2023protst} & \checkmark & & \checkmark & 87.8 & 89.4 & 66.1 & \underline{64.4} & \underline{48.8} & \underline{36.4} & \underline{48.0} & 32.8 & \cellcolor{gray!25} \underline{62.7} & \cellcolor{gray!25} \underline{55.8}\\
    \midrule
    AMMA-symmetric (ours) & \checkmark & \checkmark & \checkmark & 71.6 & 74.9 & 52.0 & 52.3 & \underline{48.8} & 35.1 & 35.5 & 24.3 & \cellcolor{gray!25} 52.0 & \cellcolor{gray!25} 46.7\\
    AMMA-contrastive (ours) & \checkmark & \checkmark & \checkmark & 87.7 & \underline{89.5} & 65.2 & 61.3 & 44.3 & 28.2 & 28.2 & 17.3 & \cellcolor{gray!25} 56.4 & \cellcolor{gray!25} 49.1\\
    \midrule
    AMMA (ours) & \checkmark & \checkmark & \checkmark & \underline{88.7} & \textbf{89.8} & \underline{67.3} & \textbf{65.5} & \textbf{49.8} & \textbf{36.9} & 46.9 & \textbf{33.6} & \cellcolor{gray!25} \textbf{63.2} & \cellcolor{gray!25} \textbf{56.5}\\
    \bottomrule
    \end{tabular}}
    \label{tab:main}
    \vspace{-0.27in}
\end{table*}

\vspace{-0.1in}

\subsection{Protein Function Prediction}
\vspace{-0.1in}
As shown in Table~\ref{tab:main}, AMMA outperforms all baselines on all tasks in terms of AUPR and on two out of four tasks in terms of $\text{F}_\text{max}$. This demonstrates that the proposed pretraining scheme with AMMA is highly effective in learning high-quality protein representations that can be universally used in downstream tasks to improve performance. Specifically, AMMA largely outperforms its sequence encoder ESM-1b and its structure encoder GearNet in terms of average score, showing that utilizing protein representations that integrate information from multiple modalities benefits prediction performance. On the other hand, AMMA shows better or comparable results to ProtST, a model that use significantly more parameters and training resources than AMMA. While ProtST has 650M parameters and takes 205 hours to pretrain, AMMA has 111M parameters and takes only 120 hours to train on fewer GPUs as shown in Section~\ref{sec:time}. This result demonstrates that AMMA learns comprehensive protein representations in an efficient and effective way to improve the performance of a wide range of downstream tasks.

\vspace{-0.1in}
\subsection{Improving Performance with Unpaired Data}
\vspace{-0.1in}
\begin{wrapfigure}[6]{r}{0.45\textwidth}
  \vspace{-0.15in}
  \begin{minipage}{0.44\textwidth}
    \captionof{table}{\small \textbf{EC/GO results of 15 epochs with extra unpaired data.}}
    \vspace{-0.1in}
    \centering
    \scalebox{0.6}{
    \renewcommand{\tabcolsep}{2.5mm}
    \begin{tabular}{cc|ccccc}
    \toprule
    \multicolumn{2}{c|}{Data} & \multicolumn{2}{c}{EC} & \multicolumn{2}{c}{GO-MF}& \multirow{2.5}{*}{Average} \\
    \cmidrule(l{2pt}r{2pt}){1-2}
    \cmidrule(l{2pt}r{2pt}){3-4}
    \cmidrule(l{2pt}r{2pt}){5-6}
    Paired & Unpaired & $\text{F}_\text{max}$ & AUPR & $\text{F}_\text{max}$ & AUPR \\
    \midrule
    120k&0k&88.1&89.7&66.4&\textbf{64.6}&\cellcolor{gray!25} 77.2\\
    120k&50k&\textbf{88.2}&\textbf{90.4}&\textbf{66.9}&\textbf{64.6}&\cellcolor{gray!25} \textbf{77.5}\\
    \bottomrule
    \end{tabular}}
    \label{tab:unpaired}
  \end{minipage}
\end{wrapfigure}

One of the biggest limitations of contrastive learning approach is its inability to leverage abundant unpaired data, i.e., data without all modality information. On the contrary, the pretraining strategy using AMMA can be flexibly applied to unpaired data under the masked autoencoder framework. Since there is much more data that only partially has the sequence-structure-function triplet, this is a huge advantage and can be leveraged to further improve the performance of AMMA.
We construct an unpaired dataset of 50k with 25k sequence-structure pairs randomly selected from AlphaFoldDB and 25k sequence-function pairs randomly selected from ProtDescribe and further train AMMA on them. As shown in Table~\ref{tab:unpaired}, we found that further pretraining AMMA on the unpaired data improves prediction performance, demonstrating the scalability of AMMA. The detailed experimental setting is provided in Section~\ref{sec:unpair_detail}.

\vspace{-0.1in}
\subsection{Ablation Studies and Qualitative Analysis}
\vspace{-0.1in}
\paragraph{Effect of the Asymmetric Decoders}
To capture the asymmetric relationship between protein sequence, structure, and function, AMMA employs asymmetric decoding, where the structure decoder predicts structural features by considering sequence and function, and the function decoder predicts functional features by considering sequence and structure. To examine the effect of the asymmetric decoding, we compare AMMA to \textbf{AMMA-symmetric}, a variant of AMMA that uses symmetric decoding. The sequence decoder, structure decoder, and function decoder of AMMA-symmetric take sequence, structure, and function inputs, respectively, and predict its own features, i.e., sequence, structure, and function features, respectively, without reference to other modalities. 

\vspace{-0.05in}
\begin{wrapfigure}[12]{r}{0.43\textwidth}
  \centering
  \vspace{-0.15in}
  \includegraphics[width=0.9\linewidth]{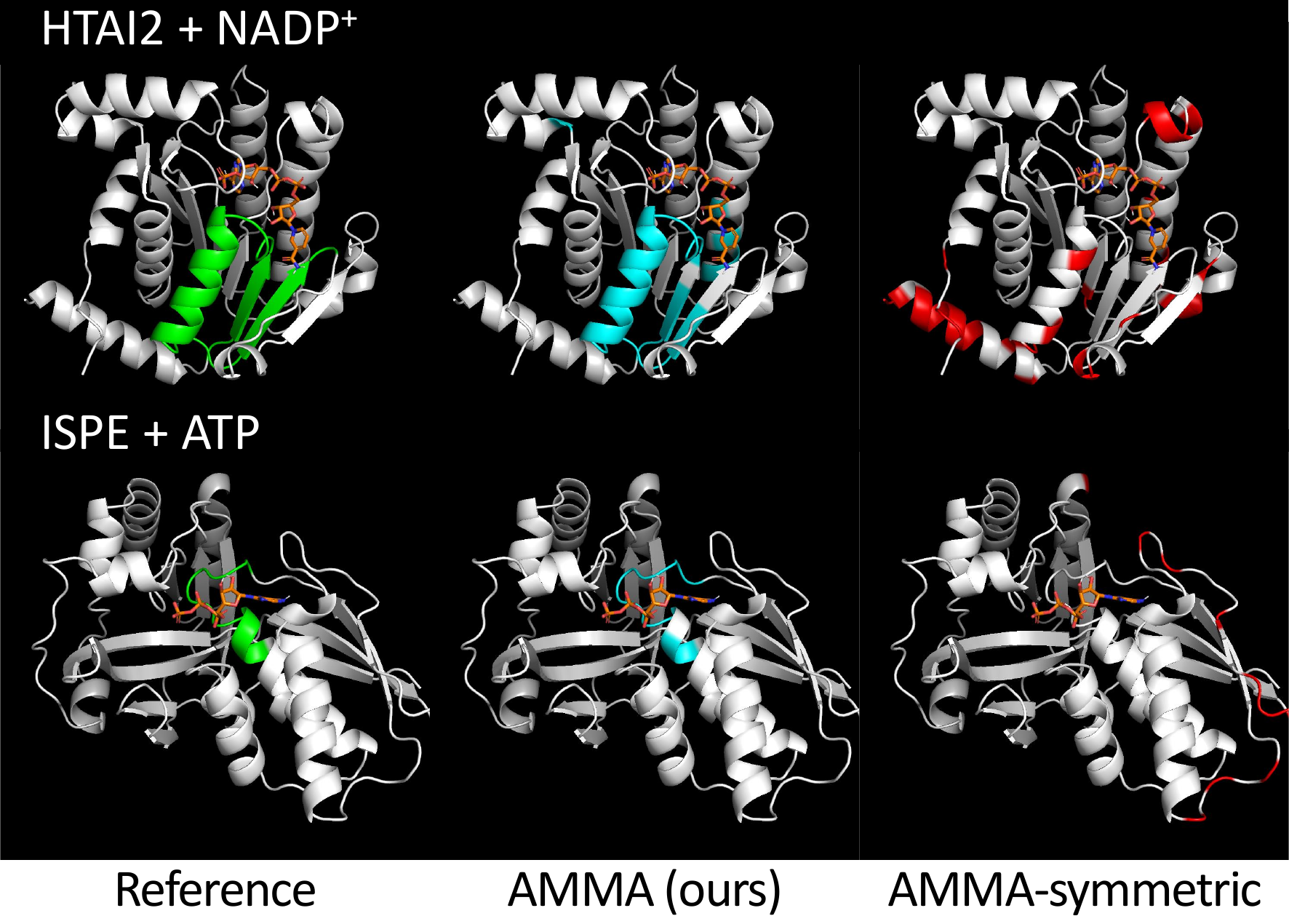}
  \vspace{-0.13in}
  \caption{\small \textbf{Visualization of highly attended residues in a functional context.}}
  \label{fig:attention_map}
\end{wrapfigure}

As shown in Table~\ref{tab:main}, AMMA outperforms AMMA-symmetric by a large margin, demonstrating that the asymmetric design of decoders greatly benefits performance as it allows the multi-modal encoder to encode a more comprehensive protein representation by considering the asymmetric relationship between the protein modalities. 

To further investigate the effect of the asymmetric decoders and understand the inter-modal relationship at the residue level, we visualize the residues with the highest function-to-residue attention values in AMMA and AMMA-symmetric. As shown in Figure~\ref{fig:attention_map}, residues heavily attended by functional tokens correspond to actual protein-ligand binding regions in AMMA, while heavily attended residues in AMMA-symmetric do not correlate with the actual interaction region. This demonstrates that the asymmetric design of decoders helps AMMA understand the relationship between functional context and residual information. Furthermore, this also suggests that AMMA could be utilized to find the active region of a protein, another important research problem with various applications. We provide the experimental details in Section~\ref{sec:attn_vis}.

\vspace{-0.1in}
\paragraph{Comparison with Contrastive Learning}
We also compare AMMA to contrastive learning, a popular pretraining strategy for learning relationships between multiple modalities. Specifically, we construct \textbf{AMMA-contrastive}, a model that uses a similar architecture to AMMA but uses contrastive learning to obtain multi-modal protein representations. 

As shown in Table~\ref{tab:main}, AMMA largely outperforms AMMA-contrastive, indicating that the proposed strategy to utilize a unified multi-modal encoder and asymmetric decoders is much more effective than contrastive learning in integrating the multi-modal information of proteins. Compared to its uni-modal sequence encoder ESM-1b, AMMA-contrastive shows better EC prediction performance but worse GO prediction performance, suggesting that contrastive learning is not beneficial to all tasks, while AMMA learns high-quality protein representations that are universally applicable to a variety of downstream tasks. 
\begin{figure}[t]
  \centering
  \begin{subfigure}[b]{0.48\textwidth}
    \centering
    \includegraphics[width=0.85\linewidth]{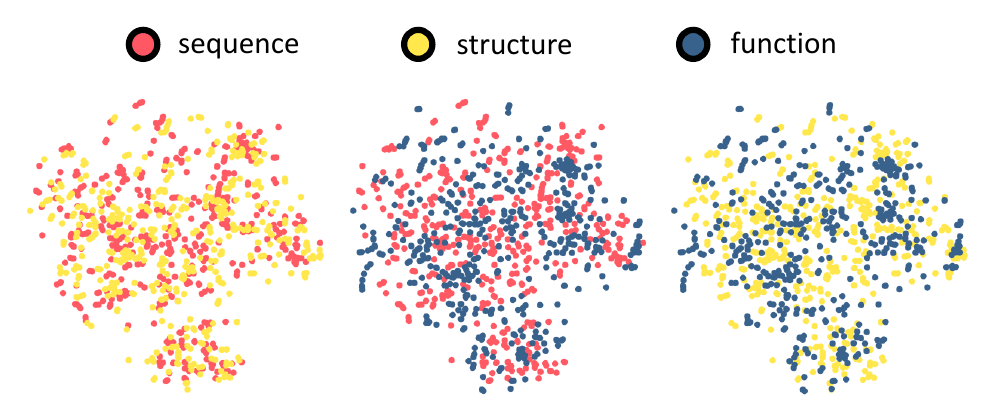}
  \vspace{-0.15in}
    \caption{AMMA}
    \label{fig:tsne_2}
  \end{subfigure}
  \hfill
  \begin{subfigure}[b]{0.48\textwidth}
    \centering
    \includegraphics[width=0.85\linewidth]{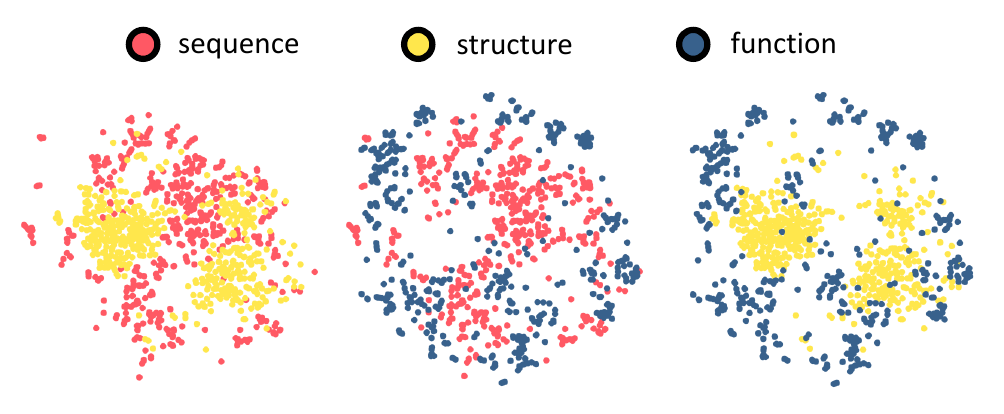}
    \vspace{-0.15in}
    \caption{AMMA-contrastive}
    \label{fig:tsne_3}
  \end{subfigure}
  \vspace{-0.1in}
  \caption{\small t-SNE visualization of the three protein modalities after (a) AMMA training and (b) contrastive learning. Sequence, structure, and function features are well-aligned after training with AMMA while contrastive learning fails to align the three modalities in a balanced manner. Details are provided in Section~\ref{sec:tsne_detail}.}
  \vspace{-0.26in}
\end{figure}

In addition, we provide t-SNE visualization of the uni-modal latent features obtained after training with AMMA and contrastive learning in Figure~\ref{fig:tsne_2} and~\ref{fig:tsne_3}, respectively. While the features of the difference modalities after training with AMMA are well aligned, those after contrastive learning are not aligned in a balanced way, because AMMA-contrastive only utilizes structure and function features to guide sequence features and does not uniformly fuse the multi-modal information. This results can also quantitatively reconfirmed in Table~\ref{tab:similarity}, indicating that AMMA uniformly integrates information of multiple modalities while the contrastive learning approach cannot.


\vspace{-0.1in}
\begin{wraptable}{r}{0.5\columnwidth}
    \vspace{-0.15in}
    \caption{\small \textbf{Experimental results with different $\alpha$ for Dirichlet sampling.} The pretraining is conducted using a 22k dataset, a random subset of the 120k dataset.}
    \vspace{-0.1in}
    \centering
    \resizebox{\linewidth}{!}{
    \renewcommand{\tabcolsep}{2.5mm}
    \begin{tabular}{ccc|ccccc}
    \toprule
    \multicolumn{3}{c|}{Ratio} & \multicolumn{2}{c}{EC} & \multicolumn{2}{c}{GO-MF} & \multirow{2.5}{*}{Average} \\
    \cmidrule(l{2pt}r{2pt}){1-3}
    \cmidrule(l{2pt}r{2pt}){4-5}
    \cmidrule(l{2pt}r{2pt}){6-7}
   $\alpha_{\texttt{seq}}$ & $\alpha_{\texttt{str}}$ & $\alpha_{\texttt{func}}$ & $\text{F}_\text{max}$ & AUPR & $\text{F}_\text{max}$ & AUPR & \\
    \midrule
    1 & 1 & 1 & 84.6 & 87.2 & 66.4 & 64.8 & \cellcolor{gray!25} 75.8\\
    1 & 2 & 2 & 87.7 & 89.8 & 66.4 & 64.2 & \cellcolor{gray!25} \textbf{77.0}\\
    2 & 1 & 1 & 73.0 & 75.9 & 65.1 & 63.2 & \cellcolor{gray!25} 69.3\\
    2 & 1 & 2 & 86.7 & 89.2 & 65.9 & 64.7 & \cellcolor{gray!25} 76.6\\
    2 & 2 & 1 & 87.9 & 89.5 & 52.4 & 53.6 & \cellcolor{gray!25} 70.9\\
    \bottomrule
    \end{tabular}}
    \label{tab:ablation}
    \vspace{-0.2in}
\end{wraptable}

\paragraph{Effect of the Masking Ratio}
We examine the effect of the $\alpha$ value used for Dirichlet sampling to sample the masking ratio. As shown in Table~\ref{tab:ablation}, the choice of $\alpha$ value has a large impact on the final performance of AMMA. Note that the higher the $\alpha$ value, the less masking is applied and the more tokens are preserved. We can observe a trend that preserving fewer sequence features favors model performance. This is due to the two reasons. First, as we saw in the previous paragraph, auxiliary structure or function features are important for successful pretraining of AMMA, so preserving them over less aligned sequence features is beneficial to performance. This can be seen by the fact that the third row performs the worst due to minimal auxiliary features, while the fourth and fifth rows perform better by utilizing more auxiliary information. Second, masking more tokens in a sequence feature make the sequence feature more robust because it must contain a wealth of information in its small subset. As a result, the best performance is achieved by setting $\alpha_\texttt{seq}$, $\alpha_\texttt{str}$, and $\alpha_\texttt{func}$ to 1, 2, and 2, respectively, which are the values used in the main experiment (Table~\ref{tab:main}).


\vspace{-0.1in}
\section{Conclusion}
\vspace{-0.1in}
In this paper, we proposed AMMA to address the problem of multi-modal protein representation learning that considers three core protein modalities: sequence, structure, and function. AMMA utilizes a unified multi-modal encoder and asymmetric decoders to capture the asymmetric relationship between the protein modalities, resulting in high-quality, comprehensive multi-modal protein representations. Through various experiments, AMMA demonstrated its effectiveness on protein-related downstream tasks with much less pretraining data. We believe that AMMA can be applied to improve our understanding of proteins by providing a comprehensive view of their properties, and we expect AMMA to spawn many interesting future studies.

\bibliography{references}
\bibliographystyle{mlgenx2024}

\clearpage
\onecolumn
\appendix
\begin{center}{\bf {\LARGE Appendix}}\end{center}

\section{Related Works~\label{sec:related_works}}
\subsection{Uni-modal Protein Representation Learning}
For learning protein representations using a single modality, the most basic and widely used is the sequence of the protein, i.e., a linear chain of amino acid residues. Many previous sequence-based protein representation learning methods have utilized language modeling techniques such as masked language modeling (MLM)~\citep{devlin2018bert}. ProtBERT~\citep{elnaggar2021prottrans} used BERT~\citep{devlin2018bert} to reconstruct missing amino acid residues. ESM-1b~\citep{rives2021biological} conducted masked language modeling (MLM) unsupervised learning on 250M protein sequences. ESM-1v~\citep{meier2021language} focused on capturing the effect of variations in sequence on function. ESM-2~\citep{lin2022language} proposed a language model with the larger 15B parameters than ESM-1b. \citet{heinzinger2022contrastive} proposed to obtain optimized sequence embeddings for the CATH protein structure hierarchy~\citep{orengo1997cath} by using contrastive learning to maximize the distance between sequences from different CATH classes and minimize the distance for those within the same class. Recently, the importance of structural information in learning protein representations has came to the fore and recent works have been proposed to exploit structural features in protein 3D geometry. \citet{hermosilla2022contrastive} and \citet{zhang2022protein} proposed to learn geometric features through contrastive learning between substructures of a given protein.
While these sequence- or structure-based methods can capture properties of proteins thanks to the vast amount of protein data available, they are limited to learning a single modality of proteins and cannot obtain comprehensive protein representations.

\vspace{-0.05in}
\subsection{Multi-modal Protein Representation Learning}
\vspace{-0.05in}
\paragraph{Contrastive Learning-based Multi-modal Learning}
As a means to overcome the limitations of uni-modal protein representation learning, protein representation learning using multiple modalities has gained traction. Most previous studies have adopted a contrastive learning approach to capture the relationship between modalities. \citet{zhang2022ontoprotein} employed knowledge-aware negative sampling to identify negative instances, enabling contrastive learning across proteins. \citet{xu2023protst} conducted contrastive learning between protein sequence and functional description. \citet{zhang2024pepharmony} utilized ESM~\citep{rives2021biological} and GearNet~\citep{zhang2022protein} to encode sequence and structure data, respectively, and performed contrastive learning between the encoded sequence and structure features. However, contrastive learning may not be an optimal choice for multi-modal protein representation learning, as it essentially focuses on learning improved representations of a single modality using other modal information and thus cannot yield balanced multi-modal representations.

\vspace{-0.15in}
\paragraph{Other Multi-modal Learning Approaches}
Apart from contrastive learning, there are other approaches to multi-modal protein representation learning. \citet{chen2023structure} incorporated protein structural information by learning to predict residue distance or dihedral angle. This work also proposed to maximize the mutual information between the sequential representation and structural representation. \citet{su2023saprot} proposed using structure-aware tokens extracted from FoldSeek~\citep{van2022foldseek} to train a token-based ESM~\citep{lin2022language} backbone. These methods showed superior results as they consider the structure of proteins. However, the method of \citet{chen2023structure} has the limitation that it only uses sequence information as an auxiliary to guide and enhance structural representation. The method of \citet{su2023saprot} uses 20 different structural tokens, which restricts diversity when encoding structural information. Moreover, these methods do not consider another important modality for proteins, the function description, and are therefore suboptimal.

\section{Experimental Details}
\subsection{Details on Pretraining~\label{sec:pretrain_detail}}

\paragraph{Details} For pretraining, we set the batch size as 4 and the number of training epochs as 10. We used the AdamW~\citep{adamw} optimizer with a learning rate of $1\times10^{-4}$ and a weight decay of $0.05$. The pretraining scheduler was StepLR with gamma $0.5$.

\paragraph{Dataset} For pretraining, we built a dataset of 120k sequence, structure, and function triplets. We extracted common proteins from the AlphaFold v2 dataset~\citep{varadi2022alphafold} of 440k sequence and structure pairs and the ProtDescribe~\citep{xu2023protst} dataset of 553k sequence and function description pairs. We used the preprocessed sequences of AlphaFold using the torchdrug~\citep{zhu2022torchdrug} library when extracting the common proteins.

\subsection{Details on Protein Function Prediction~\label{sec:protein_function_prediction}}
\paragraph{Downstream Tasks}
We evaluated the performance of our model on two standard downstream tasks following \citet{gligorijevic2021structure}: Enzyme Commission (EC) number prediction and Gene Ontology (GO) term prediction. The GO benchmark has three sub-tasks, i.e., the tasks to predict Molecular Function (GO-MF), Cellular Component (GO-CC), and Biological Process (GO-BP), respectively.

\paragraph{Supervised Finetuning}
To finetune AMMA, sequence and structural features extracted from uni-modal encoders, respectively, were concatenated and passed to the multi-modal encoder without masking. Following \citet{devlin2018bert}, we use the latent feature corresponding to first token as the multi-modal representation. We add two fully connected (FC) layers after the multi-modal encoder and use cross entropy loss to finetune the sequence encoder, multi-modal encoder, and two FC layers together.

We set the batch size to 1 on EC, GO-MF, and GO-CC and 4 on GO-BP. We trained for 100 epochs on EC, and 50 epochs on GO. For EC and GO-BP, we used the AdamW~\citep{adamw} optimizer with a learning rate of 
$1.0 \times 10^{-5}$ for $ \textsc{Enc}_{\texttt{multi}}$, $\textsc{Enc}_\texttt{seq}$, and $\textsc{Proj}_{\texttt{str}}$, and $1.0 \times 10^{-4}$ for the Multi-Layer Perceptron (MLP) classifier. For GO-MF and GO-CC, we used the Adam optimizer with a learning rate of 
$1.0 \times 10^{-5}$ for $ \textsc{Enc}_{\texttt{multi}}$, $\textsc{Enc}_\texttt{seq}$, and $\textsc{Proj}_{\texttt{str}}$, and $1.0 \times 10^{-4}$ for the Multi-Layer Perceptron (MLP) classifier. We used the ExponentialLR scheduler with a gamma value of $0.95$.

\paragraph{Baselines}
We compared AMMA with five protein representation learning baselines. \textbf{ESM-1b}~\citep{rives2021biological} used masked language modeling (MLM) to learn protein representations from a large sequence database. \textbf{OntoProtein}~\citep{zhang2022ontoprotein} incorporated knowledge graphs (KGs) to enhance protein sequence embeddings with biological knowledge facts. \textbf{GearNet}~\citep{zhang2022protein} leveraged multiview contrastive learning to train the structural encoder. \textbf{SaProt}~\citep{su2023saprot} used structure-aware tokens to incorporate structure information to ESM~\citep{lin2022language}. \textbf{ProtST}~\citep{xu2023protst} used contrastive learning to align sequence information from a protein language model with functional information from a biomedical language model.

\paragraph{Finetuning Dataset}
We tested on the Enzyme Commission (EC) and Gene Ontology (GO) in downstream tasks. We used a 95\% sequence identity cutoff for both EC and GO, following GearNet~\citep{zhang2022protein}. The dataset size of each dataset is shown in Table~\ref{tab:finetuning}.

\begin{table}[H]
    \caption{\small\textbf{The size of the finetuning datasets.}}
    \vspace{-0.1in}
    \centering
    \resizebox{0.45\columnwidth}{!}{
    \begin{tabular}{c|ccc}
    \toprule
    & \# Train & \# Valid & \# Test \\
    \midrule
    Enzyme Commission (EC) & 15,035 & 1,665 & 1,840\\
    Gene Ontology (GO) & 27,581 & 3,061 & 2,991\\
    \bottomrule
    \end{tabular}}
    \label{tab:finetuning}
\end{table}

\paragraph{Evaluation Metrics}
To quantify the effectiveness of protein representation learning methods, we used two commonly used metrics: the protein-centric maximum F-score ($\text{F}_\text{max}$) and pair-centric area under precision-recall curve (AUPR) that are implemented in torchdrug~\citep{zhu2022torchdrug}.

The $\text{F}_\text{max}$ is the protein-centric maximum F-score. $t \in [0,1]$ denotes a decision threshold for a target protein $i$, and we calculated precision and recall as follows:
\begin{align}
\text{precision}_i(t) &= \frac{\sum_{f} \mathbbm{1}[f \in P_i(t) \cap T_i]}{\sum_{f} \mathbbm{1}[f \in P_i(t)]}, \\
\text{recall}_i(t) &= \frac{\sum_{f} \mathbbm{1}[f \in P_i(t) \cap T_i]}{\sum_{f} \mathbbm{1}[f \in T_i]},
\end{align}
where $f$ indicates a functional term in EC or GO ontology. For protein $i$, $T_i$ denotes the set comprising all experimentally validated functional terms for the protein. $P_i(t)$ is the set of predicted functional terms for protein i, each with a score at least threshold $t$. $\mathbbm{1}[\cdot]$ denotes an indicator function.

The average precision and recall with threshold $t$ for all proteins are defined as follows:
\begin{align}
\text{precision}(t) &= \frac{1}{M(t)} \sum_{i} \text{precision}_i(t), \\
\text{recall}(t) &= \frac{1}{N} \sum_{i} \text{recall}_i(t),
\end{align}
where $N$ is the number of proteins, and $M(t)$ represents the count of proteins that have at least one predicted function exceeding the threshold $t$.

$\text{F}_{\text{max}}$ is calculated as follows:
\begin{align}
\text{F}_{\text{max}} = \max_t \left\{ \frac{2 \cdot \text{precision}(t) \cdot \text{recall}(t)}{\text{precision}(t) + \text{recall}(t)} \right\}.
\end{align}

The second metric, AUPR is pair-centric area under precision-recall curve which calculate the average precision score over all protein-function pairs.


\subsection{Number of Pretraining Data, Parameters and Training Time~\label{sec:time}}
We compare the number of pretraining data, number of parameter, and training time required for pretraining ProtST~\citep{xu2023protst} and AMMA. ProtST was pretrained on 553k dataset, while AMMA was pretrained on 120k dataset. AMMA has 111M parameters, while ProtST has 675M parameters. ProtST takes 205 hours of pretraining using 4 Tesla V100 GPUs, while AMMA takes 120 hours using 2 NVIDIA RTX 3090 GPUs. AMMA requires less amount of pretraining data, less number of parameters, and shorter training time, while achieving better performance. This shows that AMMA is a very efficient and powerful model with much less pretraining data, parameters, and training time.

\subsection{Details on the t-SNE Visualization~\label{sec:tsne_detail}}
In Figure~\ref{fig:tsne}, Figure~\ref{fig:tsne_2}, and Figure~\ref{fig:tsne_3}, we visualized representation space of encoders using t-SNE. In this section, we describe in more detail how we performed t-SNE.

To visualize t-SNE in Figure~\ref{fig:tsne}, a subset of 1,000 paired data points was randomly selected from the larger 120k pretraining dataset for visualization purposes. Prior to t-SNE visualization, Principal Component Analysis (PCA) was applied to reduce the dimension of the outputs from each modality encoder, namely ESM-1b, GearNet, and PubMedBERT, to 100, and the resulting latent representations were then visualized in 2D using t-SNE. The t-SNE algorithm was executed for 2,500 iterations with a perplexity setting of 200. To incorporate real data insights, two specific instances, representing the proteins 30S ribosomal S13 and 50S ribosomal L22, were included. Furthermore, the 3D structures were visualized using the PDB data available on AlphaFold\footnote{\url{https://alphafold.ebi.ac.uk}}.

In Figure~\ref{fig:tsne_2}, 500 paired data points were randomly selected for visualization from the 120k pretraining dataset. These paired data were forwarded to our multi-modal encoder without masking,  resulting in $Z'_{\texttt{seq},0} \in \mathbb{R}^{L \times 512}$, $Z'_{\texttt{str},0} \in \mathbb{R}^{L \times 512}$, $Z'_{\texttt{func},0} \in \mathbb{R}^{L' \times 512}$ similar to Eq.~\eqref{eq:decoder}. We then average theses vectors over the length dimension to make $z''_{\texttt{seq},0} \in \mathbb{R}^{512}$, $z''_{\texttt{str},0} \in \mathbb{R}^{512}$, $z''_{\texttt{func},0} \in \mathbb{R}^{512}$. Before performing principle component analysis (PCA), we scaled the latent of each modality based on its minimum and maximum values for normalization, and then performed PCA to reduce the dimensionality to 500 components for each $z''$.
Finally, we concatenated the latents and applied t-SNE to visualize into 2D space.

For Figure~\ref{fig:tsne_3}, we used $z_\texttt{seq}$, $z_\texttt{str}$, $z_\texttt{func}$ from \eqref{eq:con} instead of $z''_{\texttt{seq}}$, $z''_{\texttt{str}}$, $z''_{\texttt{func}}$ in the above paragraph. 

\subsection{Details on the Cosine Similarity Calculation~\label{sec:csim_detail}}
To compute the cosine similarity in Table~\ref{tab:similarity}, we first normalized each latent to the L2 norm. We will denote sequence latent, structure latent, function latent used for the cosine similarity calculation process as $z_\texttt{seq}$, $z_\texttt{str}$, $z_\texttt{func}$. When computing the first column of Table~\ref{tab:similarity}, $X'_\texttt{seq}$, $X'_\texttt{str}$, $X'_\texttt{func}$ from \ref{sec:encoder} worked as $z_\texttt{seq}$, $z_\texttt{str}$, $z_\texttt{func}$. When computing the second column, $z_\texttt{seq}$, $z_\texttt{str}$, $z_\texttt{func}$ from \ref{sec:cl} worked as $z_\texttt{seq}$, $z_\texttt{str}$, $z_\texttt{func}$. When computing the third column, $Z'_{\texttt{seq},0}$, $Z'_{\texttt{str},0}$, $Z'_{\texttt{func},0}$ from Eq.~\eqref{eq:decoder} worked as $z_\texttt{seq}$, $z_\texttt{str}$, $z_\texttt{func}$. We then computed the relation matrix (Relation) for each latent by performing matrix multiplication with its transposed counterparts as follows:
\begin{align}
\begin{split}
    \text{Relation}_\texttt{seq} &= z_\texttt{seq} \cdot z_\texttt{seq}^T,\\
    \text{Relation}_\texttt{str} &= z_\texttt{str} \cdot z_\texttt{str}^T,\\
    \text{Relation}_\texttt{func} &= z_\texttt{func} \cdot z_\texttt{func}^T.\\
\end{split}
\end{align}

The resulting relation matrices contains the interrelationships between the various protein features within the modalities. We then calculated cosine similarity (CosineSim) to capture the relationships between the modalities as follows:
\begin{align}
\begin{split}
    \text{CosineSim}_\texttt{seq-str} &= \frac{\text{Relation}_\texttt{seq} \cdot \text{Relation}_\texttt{str}}{\max(\|\text{Relation}_\texttt{seq}\|_2 \cdot \|\text{Relation}_\texttt{str}\|_2, \epsilon)},\\
    \text{CosineSim}_\texttt{seq-func} &= \frac{\text{Relation}_\texttt{seq} \cdot \text{Relation}_\texttt{func}}{\max(\|\text{Relation}_\texttt{seq}\|_2 \cdot \|\text{Relation}_\texttt{func}\|_2, \epsilon)},\\
    \text{CosineSim}_\texttt{str-func} &= \frac{\text{Relation}_\texttt{str} \cdot \text{Relation}_\texttt{func}}{\max(\|\text{Relation}_\texttt{str}\|_2 \cdot \|\text{Relation}_\texttt{func}\|_2, \epsilon)}.\\
\end{split}
\end{align}

$\epsilon$ is a sufficiently small number ($10^{-8}$) to avoid division by zero. This cosine similarity measure is averaged over the training dataset to get the final value reported in Table~\ref{tab:similarity}.

\subsection{Details on Experimental Results with Unpaired Data~\label{sec:unpair_detail}}
We further trained the 120k-pretrained AMMA with the 50k dataset consists of 25k sequence-structure pairs and 25k sequence-function pairs. At each step, the batch of sequence-structure pairs and the batch of sequence-function pairs were updated simultaneously. Let us denote sequence and structure data of the sequence-structure pairs as $X_{\texttt{seq}1}'$, $X_{\texttt{str}1}'$ and the sequence and function data of the sequence-function pairs as $X_{\texttt{seq}2}'$, $X_{\texttt{func}2}'$. AMMA was trained using the following objective function:
\begin{align}
\begin{split}
    \mathcal{L}_{\texttt{seq}1} &= \text{MSE}(\hat{X}_{\texttt{seq}1}, \, X_{\texttt{seq}1}'),\\
    \mathcal{L}_{\texttt{str}1} &= \text{MSE}(\hat{X}_{\texttt{str}1}, \, X_{\texttt{str}1}'),\\
    \mathcal{L}_{\texttt{seq}2} &= \text{MSE}(\hat{X}_{\texttt{seq}2}, \, X_{\texttt{seq}2}'),\\
    \mathcal{L}_{\texttt{func}2} &= \text{MSE}(\hat{X}_{\texttt{func}2}, \, X_{\texttt{func}2}'),\\
    \mathcal{L} &= \mathcal{L}_{\texttt{seq}1} + \mathcal{L}_{\texttt{str}1} + \mathcal{L}_{\texttt{seq}2} + \mathcal{L}_{\texttt{func}2}.
\end{split}
\end{align}

\subsection{Details on Symmetric Learning~\label{sec:symm}}
AMMA-symmetric used the following decoding strategy instead of \eqref{eq:decoder}:
\begin{align}
\begin{split}
    \hat{X}_{\texttt{seq}} &= \textsc{Dec}_\texttt{seq}(Z'_{\texttt{seq},0}), \\
    \hat{X}_{\texttt{str}} &= \textsc{Dec}_\texttt{str}(Z'_{\texttt{str},1}), \\
    \hat{X}_{\texttt{func}} &= \textsc{Dec}_\texttt{func}(Z'_{\texttt{func},2}).
\end{split}
\vspace{-0.1in}
\end{align}
We provide the overall architecture of AMMA-symmetric in Section~\ref{sec:arch_AMMA_symm}.

\subsection{Details on Attention Visualization~\label{sec:attn_vis}}
To understand the inter-modal relationship, we concatenated the latent features of sequence, structure, and protein name without masking. After forwarding these features through the multi-modal encoder, we extracted the function-to-residue attention from the whole self-attention map of the last layer and averaged the attention values of each residue token with respect to the function tokens. Then, we selected residues with the highest attention values in the same number as the actual interacting residues annotated from UniProtKB~\citep{uniprot2019uniprot}. For visualization, the conformations for the protein-ligand interaction were simulated by AutoDockVina~\cite{eberhardt2021autodock}.

\subsection{Details on Contrastive Learning~\label{sec:cl}}
AMMA-contrastive first took a length-wise average over the uni-modal features $X_{\texttt{seq}}' \in \mathbb{R}^{L\times1280}$, $X_{\texttt{str}}' \in \mathbb{R}^{L\times3072}$, and $X_{\texttt{func}}' \in \mathbb{R}^{L'\times768}$ obtained in \ref{sec:encoder} to yield $X_{\texttt{seq}}'' \in \mathbb{R}^{1280}$, $X_{\texttt{str}}'' \in \mathbb{R}^{3072}$, and $X_{\texttt{func}}'' \in \mathbb{R}^{768}$. The features were then processed as follows:
\begin{align}
\begin{split}
    z'_{\texttt{seq}} &= \textsc{Proj}_{\texttt{seq}}(X_{\texttt{seq}}'') \in \mathbb{R}^{D}, \\
    z_{\texttt{seq}} &= \textsc{Enc}_{\texttt{con}}(z'_{\texttt{seq}}) \in \mathbb{R}^{D}, \\
    z_{\texttt{str}} &= \textsc{Proj}_{\texttt{str}}(X_{\texttt{str}}'') \in \mathbb{R}^{D}, \\
    z_{\texttt{func}} &= \textsc{Proj}_{\texttt{func}}(X_{\texttt{func}}'') \in \mathbb{R}^{D}.
    \label{eq:con}
\end{split}
\end{align}
Here, $\textsc{Enc}_{\texttt{con}}$ is a sequence encoder for contrastive learning. We adopted the same 8-layer Transformer as $\textsc{Enc}_\texttt{multi}$ of AMMA as $\textsc{Enc}_{\texttt{con}}$. By utilizing the following contrastive loss that aligns sequence-structure and sequence-function, AMMA-contrastive was trained to guide sequence features with structure and function information:
\begin{align}
\begin{split}
    \mathcal{L}_{\texttt{seq}2\texttt{str}} &= \mathcal{L}_{\text{con}}(z_{\texttt{seq}}, \, z_{\texttt{str}}), \\
    \mathcal{L}_{\texttt{seq}2\texttt{func}} &= \mathcal{L}_{\text{con}}(z_{\texttt{seq}}, \, z_{\texttt{func}}), \\
    \mathcal{L}_{\text{reg}} &= \mathcal{L}_{\text{con}}(z_{\texttt{seq}}, \, z'_{\texttt{seq}}), \\
    \mathcal{L} &= \mathcal{L}_{\texttt{seq}2\texttt{str}} + \mathcal{L}_{\texttt{seq}2\texttt{func}} + \mathcal{L}_{\text{reg}},
\end{split}
\end{align}
where $\mathcal{L}_\text{reg}$ was for regularization and contrastive loss $\mathcal{L}_\text{con}$ was defined as follows:
\vspace{-0.1in}
\begin{align}
    \mathcal{L}_{\text{con}}(z_P, z_Q) = -\frac{1}{2N}\sum^N_{i=1}\Big(&\mathrm{log}\frac{\mathrm{exp}(z_{P,i}\cdot z_{Q,i}/\tau)}{\sum^N_{j=1}\mathrm{exp}(z_{P,i}\cdot z_{Q,j}/\tau)} \\
    +&\mathrm{log}\frac{\mathrm{exp}(z_{P,i}\cdot z_{Q,i}/\tau)}{\sum^N_{j=1}\mathrm{exp}(z_{P,j}\cdot z_{Q,i}/\tau)}\Big). \notag
\end{align}
We provide the overall architecture of AMMA-contrastive in Section~\ref{sec:arch_AMMA_cl}.

\section{Additional Experiments~\label{sec:auxiliary_exp}}

\vspace{-0.05in}
\subsection{Effect of the Auxiliary Modalities in Decoders}
In addition to sequence features, the AMMA structure decoder takes function features as inputs and the AMMA function decoder takes structure features as inputs. We examined the effect of these auxiliary inputs in Table~\ref{tab:auxiliary}. \textbf{AMMA-w/o auxiliary} is the AMMA variant that only takes sequence features as inputs, i.e., that uses the following decoding strategy instead of \eqref{eq:decoder}:
\vspace{-0.05in}
\begin{align}
\begin{split}
    \hat{X}_{\texttt{seq}} &= \textsc{Dec}_\texttt{seq}(Z'_{\texttt{seq},0}), \\
    \hat{X}_{\texttt{str}} &= \textsc{Dec}_\texttt{str}(Z'_{\texttt{seq},1}), \\
    \hat{X}_{\texttt{func}} &= \textsc{Dec}_\texttt{func}(Z'_{\texttt{seq},2}).
\end{split}
\end{align}

\begin{table}[H]
    \caption{\textbf{Experimental results without auxiliary structure or function inputs in the AMMA decoders.} The pretraining is conducted using a 22k dataset, a random subset of the 120k dataset.}
    \centering
    \resizebox{0.5\linewidth}{!}{
    \renewcommand{\tabcolsep}{2.5mm}
    \begin{tabular}{c|ccccc}
    \toprule
    \multirow{2.5}{*}{Method} & \multicolumn{2}{c}{EC} & \multicolumn{2}{c}{GO-MF}& \multirow{2.5}{*}{Average} \\
    \cmidrule(l{2pt}r{2pt}){2-3}
    \cmidrule(l{2pt}r{2pt}){4-5}
    & $\text{F}_\text{max}$ & AUPR & $\text{F}_\text{max}$ & AUPR&  \\
    \midrule
    AMMA & 87.7 & 89.8 & 66.4 & 64.2 & \cellcolor{gray!25} \textbf{77.0}\\
    AMMA-w/o auxiliary & 87.7 & 90.1 & 65.0 & 63.6 & \cellcolor{gray!25} 76.6\\
    \bottomrule
    \end{tabular}}
    \label{tab:auxiliary}
\end{table}

As shown in the table, the auxiliary inputs to the decoders were beneficial to the performance. This is because structure and function are difficult to predict based on sequence alone, making pretraining of AMMA-w/o auxiliary too challenging. As structure and function are relatively well aligned and therefore easy to predict from each other, the auxiliary inputs aid the structure and function decoders.

\section{Architecture~\label{sec:arch}}
In this section, we illustrate the architecture of our model, AMMA, and two variants of AMMA: AMMA-symmetric and AMMA-contrastive.

\subsection{AMMA~\label{sec:arch_AMMA}}
\begin{figure}[H]
    \centering
    \includegraphics[width=\textwidth]{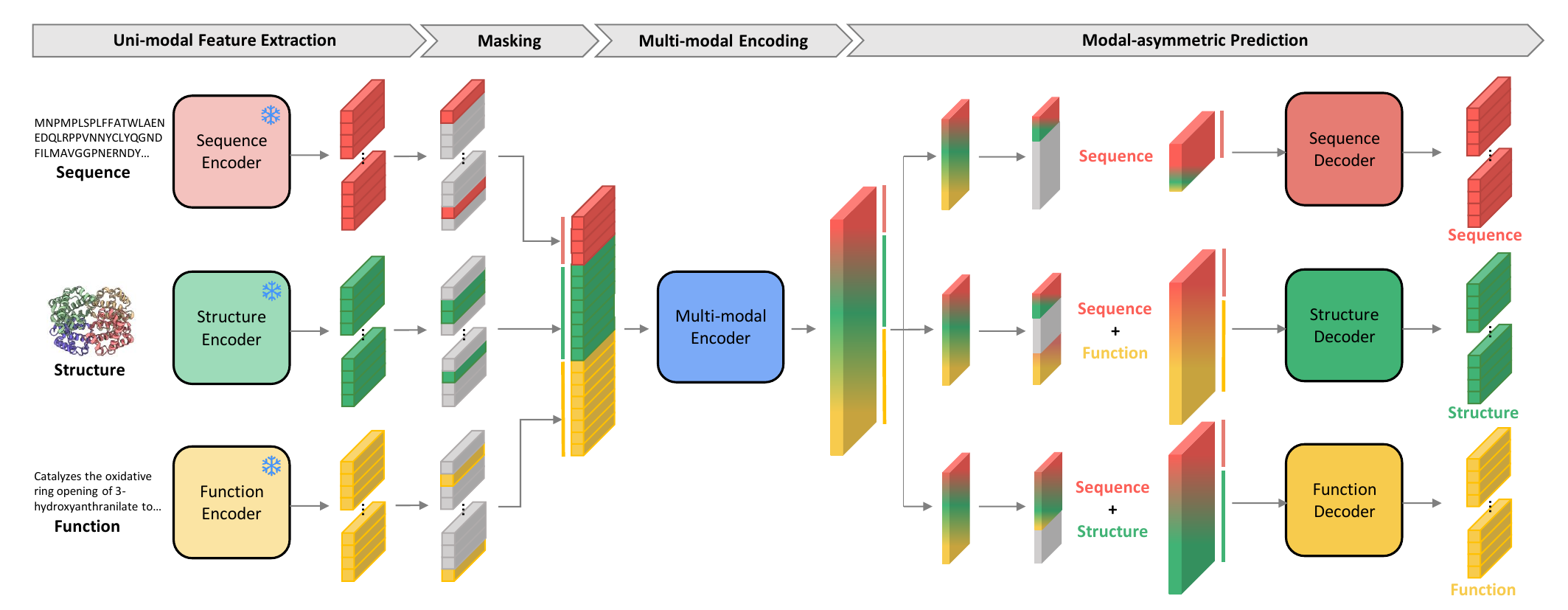}
    \caption{\textbf{The overall architecture of AMMA.}}
    \label{fig:amma}
\end{figure}

Architecture of AMMA consists of four procedures: uni-modal feature extraction, masking, multi-modal encoding, and model-asymmetric decoding as shown in Figure~\ref{fig:amma}. First, we extracted the features of each modality using a feature encoder: ESM-1b~\citep{rives2021biological}, GearNet~\citep{zhang2022protein}, PubMedBERT-abs~\citep{gu2021domain}. To be specific, the sequence encoder, ESM-1b~\citep{rives2021biological} is a model with 33 layers of Transformers with $\sim$650M parameters. The structure encoder, GearNet~\citep{zhang2022protein} is composed of 6 GearNet layers, using hidden dimension of 512. The function encoder, PubMedBERT-abs~\citep{gu2021domain} is a model based on 12 layers of BERT~\citep{devlin2018bert}. Then, we masked the latents, leaving only 160 latents in total according to the masking sampling proposed in~\ref{sec:masking}. Afterwards, we have a multi-modal encoder to fuse all modalities into a unified representation space. Finally, we performed asymmetric decoding, which reconstructs structure latent vector from sequence and function information and function latent vector from sequence and structure information, allowing AMMA to capture asymmetric sequence-structure-function relationships. The multi-modal encoder is an 8-layer transformer and the three decoders are 2-layer transformers each. 

\subsection{AMMA-symmetric~\label{sec:arch_AMMA_symm}}
\begin{figure}[H]
    \centering
    \includegraphics[width=\textwidth]{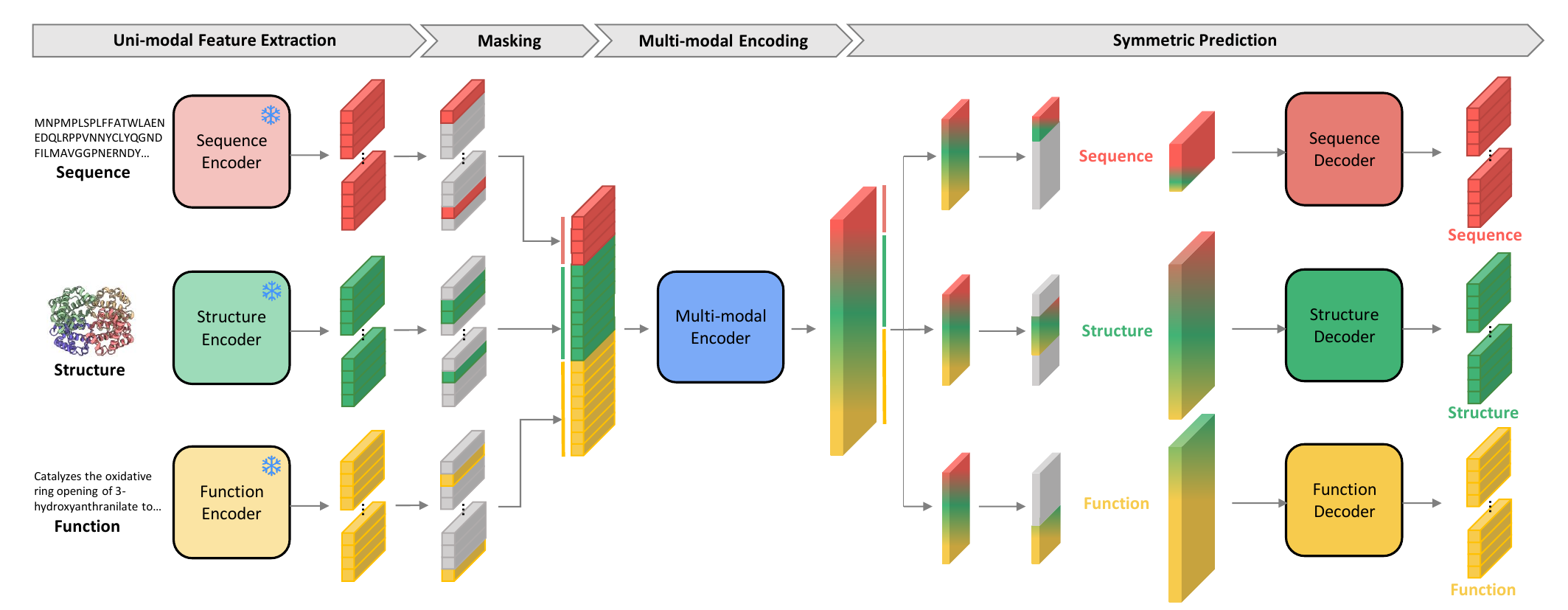}
    \caption{\textbf{The overall architecture of AMMA-symmetric.}}
    \label{fig:symm}
\end{figure}

In AMMA-symmetric, the only part that differs from the original AMMA lies in the design of decoders. Each decoder processes and predicts its corresponding modality from corresponding inputs as shown in Figure~\ref{fig:symm}. Specifically, structure latent vector is employed to structure decoder to predict the original structure latent vector. Also, function latent vector is employed to function decoder to predict the original function latent vector.

\subsection{AMMA-contrastive~\label{sec:arch_AMMA_cl}}
\begin{figure}[H]
    \centering
    \includegraphics[width=0.5\textwidth]{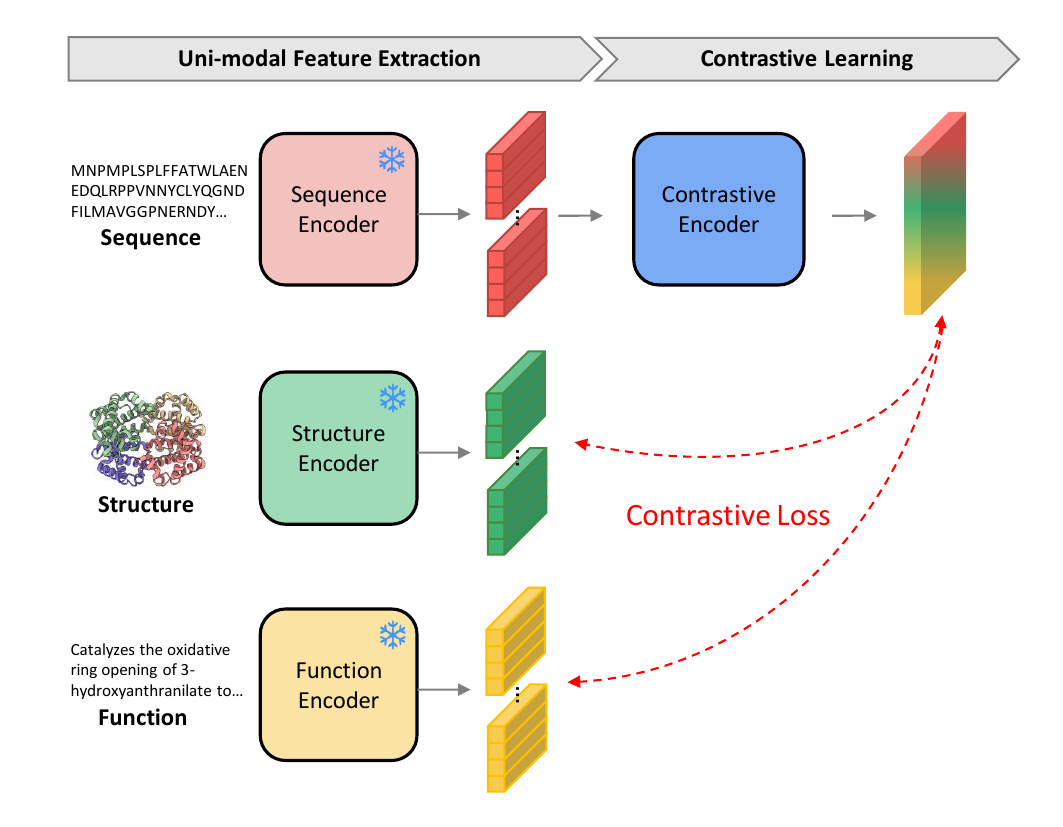}
    \caption{\textbf{The overall architecture of AMMA-contrastive.}}
    \label{fig:cl}
\end{figure}

For AMMA-contrastive, we introduced contrastive encoder with 8-layers of transformers as shown in Figure~\ref{fig:cl}. To train the AMMA-contrastive, we utilized contrastive loss between the output of contrastive encoder and structure encoder, and between the output of contrastive encoder and function encoder. We also took advantage of the regularization loss between the input and output of the contrastive encoder to regularize the output of the sequence feature extractor and the ouput of the contrastive encoder.

\end{document}